# MODELLING HYDROGEN INDUCED STRESS CORROSION CRACKING IN AUSTENITIC STAINLESS STEEL

E. I. Ogosi[1], U. B. Asim[1], M. A. Siddiq[*,1], M. E. Kartal[1]

[1] School of Engineering, University of Aberdeen, Fraser Noble Building, AB24 3UE

Aberdeen, United Kingdom

## ABSTRACT

A model has been developed which simulates the deformation of single crystal austenitic stainless steels and captures the effects of hydrogen on stress corrosion cracking. The model is based on the crystal plasticity theory which relates critical resolved shear stress to plastic strain and the strength of the crystal. We propose an analytical representation of hydrogen interactions with the material microstructure during deformation and simulate the effects hydrogen will have on void growth prior to fracture. Changes in the mechanical properties of the crystal prior to fracture are governed by the interaction of hydrogen atoms and ensembles of dislocations as the crystal plastically deforms and is based on the hydrogen enhanced localised plasticity (HELP) mechanism. The effects of hydrogen on void growth are considered by analysing the effect of hydrogen on the mechanical property of material bounding an embedded void. The model presented has been implemented numerically using the User Material (UMAT) subroutine in the finite element software (ABAQUS) and has been validated by comparing simulated results with experimental data. Influencing parameters have been varied to understand their effect and test sensitivities.

## 1. INTRODUCTION

Austenitic Stainless Steels such as AISI 304, 310 and 316 have widespread application in the nuclear, automobile, chemical, oil and gas production, refining and medical industries showing a superior strength range, ductility and corrosion resistance when compared to other types of steel [1, 2]. Austenitic Stainless Steels are however vulnerable to Stress Corrosion Cracking (SCC) in specific environmental conditions including when exposed to hydrogen. When steel is embrittled and fails by cracking due to exposure to hydrogen in the presence of stress, the failure mechanism is known as Hydrogen Induced Stress Corrosion Cracking (HISCC) [3]. Hydrogen Enhanced Localised Plasticity (HELP) is a commonly cited failure mechanism used to explain this phenomenon [4-6]. HELP suggests that hydrogen in solid solution reduces barriers to dislocation motion and increases localised deformation [7]. A manifestation of HELP in metals with low hydrogen diffusivity is strain aging and this occurs when dislocation ensembles are "pinned" by hydrogen atoms [8]. It is believed that hydrogen influences dislocation mobility by two competing mechanisms; pinning of dislocation or enhancement of dislocation mobility [6]. Robertson [6] provided evidence of increased dislocation velocity and reduced flow stress due to hydrogen. Stress-strain relationship for homogeneous solid solutions of austenitic stainless steel was studied at different temperatures and strain rates. Yield and flow stresses were observed to increase with hydrogen concentration. Yagodzinskyy and his colleagues [10] observed an increase in crystal strength and flow stress in the hydrogen charged single crystals. Their work provided experimental evidence of dislocation motion restriction due to hydrogen in single crystal austenitic stainless steel. Schebler [11] formulated a model that simulates the deformation of a single crystal face centred cubic (FCC) metal using a crystal plasticity based finite element model (FEM) and similar results to Yagodzinskyy et al [10] were obtained. The facilitation of HISCC via fracture processes of void nucleation, growth and coalescence is well established and there is experimental and theoretical evidence to support this phenomenon. S.P Lynch [12] provided a concise review of the various theories and the reader is referred to this work for more information. We use the Hydrogen Enhanced Strain Induced Vacancy (HESIV) mechanism to explain how hydrogen affects void nucleation and growth during plastic deformation [13]. HESIV proposes that hydrogen promotes strain localisation and increases vacancy formation, agglomeration to form voids, void growth and coalescence [8]. The HESIV mechanism proffers that hydrogen promotes the formation and accelerates the aggregation of vacancies introduced by plastic flow. Voids have also been observed to nucleate and grow due to intense interaction between dislocations in regions of high strain localisation [8]. Martin et al [14, 15] have previously demonstrated that the formation and extension of voids occurred along slip bands and were facilitated by the presence of hydrogen. Bullen et al [16] introduced hydrogen into interstitial sites and observed that hydrogen increased void density, but void size was reduced. Matsumoto et al [13, 17] have shown that hydrogen promotes the formation and decreases the removal of vacancies in steel during plastic deformation. Ductile fracture by vacancy condensation is considered a viable and common mechanism for transgranular failure of FCC materials including austenitic stainless steels [18] and will form the basis of modelling in this work. We extend on the above theories and model the response of a single crystal austenitic steel to applied deformation and observe the effect hydrogen has on material mechanical properties and void growth.

---

[*] Corresponding author (amir.siddiq@abdn.ac.uk)

## 2. CRYSTAL PLASTICITY THEORY

Crystal plasticity theory is extended to account for the effects of hydrogen on plastic deformation and fracture of austenitic stainless steel. We assume that the elastoplastic deformation of the crystal is driven primarily by crystal-line slip across well-defined planes and dislocation motion. The kinematics of the single crystal constitutive theory is derived from multiplicatively decomposition of the deformation gradient based on the finite strain theory [19].

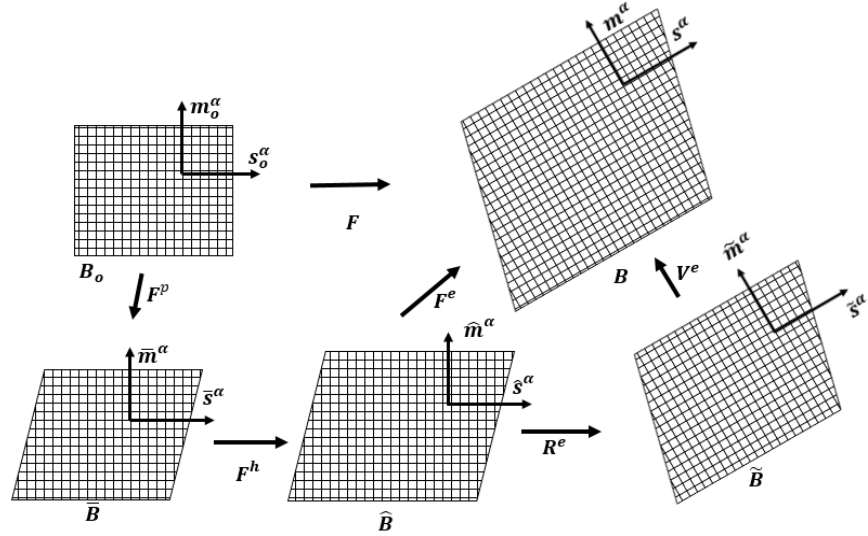

Figure 1: Representation of the multiplicative decomposition of the deformation gradient

Consider a body deformed as illustrated in Figure 1, we use vectors $\mathbf{x}^1$, $\mathbf{x}^2$, $\mathbf{x}^3$ and $\mathbf{x}^4$ to express a set of particle positions in the configurations $\mathbf{B}_o$, $\overline{\mathbf{B}}$, $\widehat{\mathbf{B}}$ and $\mathbf{B}$ respectively. $\mathbf{B}_o$ represents the original configuration and $\mathbf{B}$ represents the final deformed state at time t. Unloading elastically from $\mathbf{B}$ gives a configuration $\widehat{\mathbf{B}}$ that still has the permanent plastic deformation and the dilatational distortion induced by hydrogen. Excluding the distortional effect of hydrogen gives a second intermediate configuration, $\overline{\mathbf{B}}$. For finite strain theory, the deformation gradient matrix can be used to capture deformation histories so that $\mathbf{F}^e$, $\mathbf{F}^h$ and $\mathbf{F}^p$ are the elastic, hydrogen and plastic parts of the deformation gradient respectively. These can be expressed as follow

$$\mathbf{F} = \frac{\partial x_i^4}{\partial x_j^1} \ ; \quad \mathbf{F}^e = \frac{\partial x_i^4}{\partial x_j^3} \ ; \quad \mathbf{F}^h = \frac{\partial x_i^3}{\partial x_j^2}; \quad \mathbf{F}^p = \frac{\partial x_i^2}{\partial x_j^1} \tag{1}$$

Applying the chain rule for partial differential equations gives;

$$\frac{\partial x_i^4}{\partial x_j^1} = \frac{\partial x_i^4}{\partial x_j^3} \frac{\partial x_i^3}{\partial x_j^2} \frac{\partial x_i^2}{\partial x_j^1} \tag{2}$$

or

$$\mathbf{F} = \mathbf{F}^e \mathbf{F}^h \mathbf{F}^p \tag{3}$$

The elastoplastic formulations defined by Marin [20] are modified to include the dilatational effects of hydrogen as illustrated in figure 1;

$$\mathbf{F} = \mathbf{V}_e \mathbf{F}^*, \quad \mathbf{F}^* = \mathbf{R}^e \mathbf{F}^h \mathbf{F}^p \tag{4}$$

$\mathbf{F}$ represents the deformation gradient in the final deformed configuration $\mathbf{B}$ from an original configuration $\mathbf{B}_o$. $\mathbf{V}_e$ is the elastic stretching of the crystal. $\mathbf{R}^e$ is the rigid crystal body rotation, $\mathbf{F}^p$ represents plastic slip and is deformation gradient for the first intermediate configuration designated $\overline{\mathbf{B}}$. $\mathbf{F}^h$ represents the dilatational effect of hydrogen and is deformation gradient for the second intermediate configuration designated $\hat{B}$. The third intermediate configuration $\tilde{B}$ is obtained hypothetically by unloading elastic stretch from the final configuration $B$ through $V^{e-1}$ without excluding rotation and is represented by the deformation gradient $F^*$.

The velocity gradient in the final deformed configuration $B$ can be expressed as;

$$l = \dot{F}F^{-1} \tag{5}$$

and in the intermediate configuration $\tilde{B}$ as

$$\tilde{L} = V^{e-1}lV^e = V^{e-1}\dot{V}^e + \tilde{L}^* \tag{6}$$

By using equations (4), (5) and (6), and simplifying:

$$\tilde{L}^* = \dot{R}^e R^{eT} + R^e \hat{L}^h R^{eT} + R^e F^h \bar{L}^p F^{h-1} R^{eT} \tag{7}$$

$\tilde{L}^*$ describes plastic flow due to crystallographic slip, the dilatational effect of hydrogen and rotation of the lattice in the $\tilde{B}$ configuration. Velocity gradient due to dilatational effect of hydrogen, $\hat{L}^h$, is expressed as per the Sofronis relationship [7] as follows:

$$L^h = \dot{F}^h . F^{h-1} = \frac{d}{dt}\left(1 + \frac{(c-c_o)\lambda}{3}\right)I.\left[\left(1+\frac{(c-c_o)\lambda}{3}\right)\right]^{-1}I = \frac{1}{3}\Lambda(c)\dot{c}I, \quad \Lambda(c) = \frac{3\lambda}{3+(c-c_0)} \tag{8}$$

Here $c$ and $c_o$ are current and initial (stress-free) concentrations of hydrogen at the material point (expressed in hydrogen atoms per lattice atom). $\lambda$ is equal to $\frac{\Delta V}{V_m}$, $\Delta V$ is the volume change per atom of hydrogen introduced into solid solution and $V_m$ is the mean atomic volume of the host metal atom.

For austenitic stainless steels, bulk diffusion of hydrogen is relatively slow due to the low diffusivity of hydrogen in FCC material when compared with body centered cubic (BCC) material [9]. A comparison of hydrogen diffusivity for FCC material and experimental observation supporting constant hydrogen concentration has been discussed by Schebler [11]. Based on this, the term "$c - c_o$" tends to zero and $L^h$ is reduced to an identity matrix. The plastic deformation characteristics of the material is however altered by the presence of atomic hydrogen and a discussion of these effects will be covered in section 3.

The plastic part of velocity gradient, $\bar{L}^p$, can be defined as:

$$\bar{L}^p = \sum_{\alpha=1}^{N} \dot{\gamma}^\alpha \, \bar{s}^\alpha \otimes \bar{m}^\alpha \tag{9}$$

where $\dot{\gamma}^\alpha$ is the shear strain rate due to slip, $\bar{s}^\alpha$ and $\bar{m}^\alpha$ are the slip direction and normal to slip plane vectors respectively. After substituting these results in (7), the following relation is obtained for $\tilde{L}^*$:

$$\tilde{L}^* = \tilde{\Omega}^e + \sum_{\alpha=1}^{N}\dot{\gamma}^\alpha \, \tilde{s}^\alpha \otimes \tilde{m}^\alpha \tag{10}$$

Here, $\tilde{\Omega}^e = \dot{R}^e R^{eT}$ represents rigid body elastic spin. $\bar{s}^\alpha$ and $\bar{m}^\alpha$ are the slip direction and normal respectively. $\tilde{s}^\alpha$ is $R^e \bar{s}^\alpha$ and $\tilde{m}^\alpha$ is $R^e \bar{m}^\alpha$.

For elasticity, the Second Piola-Kirchhoff stress, $\tilde{S}$, is given as by:

$$\tilde{S} = \tilde{\mathbb{C}}^e : \tilde{E}^e \tag{11}$$

$\tilde{\mathbb{C}}^e$ is the anisotropic elasticity tensor and $\tilde{E}^e$ is the Green-Lagrange strain tensor.

We additionally decompose the velocity gradient into a symmetric, $d$, and skew, $w$, parts i.e. $l = d + w$. And in the intermediate configuration, $\tilde{B}$, the rate of deformation tensor is given as:

$$\tilde{D} = V^{eT}dV^e = \mathring{\tilde{E}}^e + \tilde{D}^* \tag{12}$$

$$\tilde{D}^* = \text{sym}(\tilde{C}^e \tilde{\Omega}^e) + \sum_{\alpha=1}^{N}\dot{\gamma}^\alpha \text{sym}(\tilde{C}^e \tilde{Z}^\alpha) \tag{13}$$

Spin of the lattice due to plastic deformation is given as:

$$\widetilde{W} = V^{eT}wV^e = skew(V^{eT}\dot{V}^e) + \widetilde{W}^* \tag{14}$$

$$\widetilde{W}^* = \text{skew}(\tilde{C}^e \tilde{\Omega}^e) + \sum_{\alpha=1}^{N}\dot{\gamma}^\alpha \text{skew}(\tilde{C}^e \tilde{Z}^\alpha) \tag{15}$$

Plastic slip evolution by shear strain rate is defined by the power law in (16). Shear strain rate $\dot{\gamma}^\alpha$ on the $\alpha^{th}$ slip system depends on the resolved shear stress and the strength of the slip system.

$$\dot{\gamma}^\alpha = \dot{\gamma}_0^\alpha \left[\frac{|\tau^\alpha|}{\kappa_s^\alpha}\right]^{\frac{1}{m}} \text{sign}(\tau^\alpha) \tag{16}$$

$\dot{\gamma}_0^\alpha$ is the reference shear strain rate on the $\alpha^{th}$ slip system, $\kappa_s^\alpha$ is the current slip system strength, $\tau^\alpha$ is resolved shear stress, $m$ is strain rate sensitivity parameter. Voce type hardening was incorporated in the model using the evolution relation in (17). The slip system is made to harden with the evolution of accumulated slip till a saturation value is reached, beyond which it deforms in a perfectly plastic manner.

$$\dot{\kappa}_s^\alpha = h_0 \left(\frac{\kappa_{s,S}^\alpha - \kappa_s^\alpha}{\kappa_{s,S}^\alpha - \kappa_{s,0}^\alpha}\right) \sum_{\alpha=1}^{N} |\dot{\gamma}^\alpha|, \quad \kappa_{s,S}^\alpha = \kappa_{s,S0}^\alpha \left[\frac{\sum_\alpha |\dot{\gamma}^\alpha|}{\dot{\gamma}_{s0}}\right]^{1\backslash m'} \tag{17}$$

$\dot{\kappa}_s^\alpha$ is the current rate of hardening, $\kappa_s^\alpha$ is the current slip system strength, $h_0$ is the reference hardening coefficient, $\kappa_{s,S}^\alpha$ is the saturation value of strength which depends on the accumulated slip $\sum_\alpha |\dot{\gamma}^\alpha|$, and its evolution is given by a power law. $\kappa_{s,0}^\alpha$ $\kappa_{s,S0}^\alpha, \dot{\gamma}_{s,0}^\alpha$ and $m'$ are the material parameters controlling the evolution of strength in the crystal. Critical resolved shear stress (CRSS) of each of the slip systems are assigned as $\kappa_s^\alpha(t=0)$ in (17).

## 3. EFFECTS OF HYDROGEN

Hydrogen atoms can be introduced into austenitic stainless steel during manufacturing, fabrication (e.g. welding), in service (transport/storage of hydrogen containing fluid, cathodic protection etc.) or generated from chemical dissociation of water due to corrosion reactions [21]. Hydrogen atoms will either reside at normal interstitial lattice sites (NILS) or in trap sites introduced to the material by plastic deformation [22]. According to Oriani's theory, hydrogen residing in these two sites are in equilibrium such that the total hydrogen concentration $C_{Total}$ is defined by the relationship below [22]:

$$C_{Total} = C_L + C_T \tag{18}$$

$C_L$ and $C_T$ represent hydrogen atoms residing in NILS and trap sites respectively. According to Fick's first law, there would be transfer of hydrogen atoms between sites if a concentration gradient exist. The proportional relationship between the hydrogen activity in traps $a_T$ and NILS $a_L$ is captured by the equilibrium constant $K_T$

$$a_T = K_T a_L \tag{19}$$

The hydrogen activity $a_i$ in each site relates to fraction of species occupancy $\theta_i$ as follows;

$$a_i = \frac{\theta_i}{1-\theta_i} \tag{20}$$

Starting from a reference of $a_i = \theta_i$ tending towards zero and combining (19) and (20) gives;

$$\frac{\theta_T}{1-\theta_T} = K_T \frac{\theta_L}{1-\theta_L} \tag{21}$$

$\theta_T$ is hydrogen occupancy of trap sites, $\theta_L$ is hydrogen occupancy of lattice sites and $K_T$ is equilibrium constant. Concentration of hydrogen residing in trapping sites $C_T$ is given as:

$$C_T = \theta_T \psi N_T \tag{22}$$

$\psi$ is the number of sites per trap and $N_T$ is the number of traps per unit lattice given as:

$$N_T = \frac{\sqrt{3}}{a_{fcc}} \rho \tag{23}$$

$a_{fcc}$ is lattice parameter for FCC metal. Evolution of bulk dislocation density $\rho$ is given as:

$$\int_0^t \dot{\rho} \, dt = (k_1 \sqrt{y}) \int_0^t /\dot{\gamma}/ dt \tag{24}$$

$\dot{\rho}$ represents incremental changes in dislocation density. $\dot{\gamma}$ represents incremental changes in strain. $k_1$ is a constant associated with immobilised dislocation and $\sqrt{y}$ is average spacing between dislocations [23]. Equation (24) considers the effects of hydrogen on material behaviour during early stages of deformation (stages 1 and 2). Experimentally, it has been observed that dislocation annihilation only becomes significant during the later stages (stage 3) of deformation [24]. We assume total hydrogen concentration is constant at each material point due to low diffusion of hydrogen in austenitic stainless steels [25]. It is important to note that although total hydrogen concentration $C_{Total}$ is constant, there is a transfer of hydrogen atoms from NILS to more energetically favourable traps created during plastic deformation. $C_T$ is expressed as a function of $C_{Total}$ and the number of traps $N_T$ by the Krom's relationship [26]

$$C_T = \frac{1}{2}\left[\frac{N_L}{K_T} + C_{Total} + N_T - \sqrt{\left(\frac{N_L}{K_T} + C_{Total} + N_T\right)^2 - 4N_T C_{Total}}\right] \quad (25)$$

$N_L$ is the number of atoms per unit NILS. We introduce two dimensionless terms to account for the changes to the material properties due to the presence of atomic hydrogen;

1) Hydrogen initial strength coefficient $H_i$ which quantifies the effect of hydrogen on initial crystal strength.
2) Hydrogen hardening coefficient $H_f$ that quantifies effect of hydrogen on strain hardening.

Initial crystal strength $\kappa_{h,0}^\alpha$ in the presence of hydrogen is given by the following expression;

$$\kappa_{h,0}^\alpha = \kappa_{s,0}^\alpha * (1 + H_i C_{initial}) \quad (26)$$

$\kappa_{s,0}^\alpha$ is crystal strength in hydrogen free condition. $C_{initial}$ is the amount of hydrogen in trapping sites before plastic deformation and relates to hydrogen uptake given by Caskey Jr [27]

$$C_{initial} = f\, C_L e^{18400/(RT)} \quad (27)$$

$f$ is a fraction of alloy atoms which are associated with a unit length of dislocation and is a function of initial dislocation density (in a strain free state). $C_L$ is the concentration of hydrogen in the lattice sites. 18400 J/mol is bonding energy for hydrogen to dislocations in austenitic stainless steels. The form of (27) is consistent with Oriani's theory (see equation 22). Hydrogen concentration in traps existing before deformation, $C_{initial}$ is shown to be directly proportional to the amount of hydrogen residing in NILS, $C_L$. The term $f$ is a material property which captures the amount of dislocations expected in austenitic stainless steel not subject to loading and is obtained empirically from testing [27].

The evolution of crystal strength given in (17) is modified as follows:

$$\dot{\kappa}_s^\alpha = h_0 \left(\frac{\kappa_{s,S}^\alpha - \kappa_s^\alpha}{\kappa_{s,S}^\alpha - \kappa_{s,0}^\alpha}\right)(1 + H_f C_T) \quad (28)$$

## 4. METHODOLOGY AND RESULTS

Finite Element Model (FEM) simulations have been performed using ABAQUS/Standard analysis [28]. Three-dimensional RVE models were constructed and meshed using the ABAQUS/CAE function with reduced-integrated, first-order linear brick elements (C3D8R). Load was applied to the RVE using displacement controlled analysis. Material response relationships and equations presented in sections 2 and 3 have been implemented numerically using the User Material (UMAT) function in ABAQUS. FEM simulations were performed by replicating the case of an austenitic stainless steel single crystal subjected to uniaxial tension oriented for multi-slip with displacement control loading applied. Tensile tests performed on the RVE specimen were done by applying identical conditions and parameters used in the real world experiments. Comparisons have then been made between simulated results and experimental data for model validation. Experimental tensile tests referenced, were performed on AISI316LN austenitic stainless steel single crystals of nominal chemical composition 18Cr-12Ni-2Mo alloyed with 0.5 wt. % of nitrogen by Yagodzinskyy et al [8]. Samples were cut parallel to the (110) crystal plane and tensile loaded in the <001> direction at a strain rate of 8 x $10^{-4}s^{-1}$ with atomic hydrogen content of 0.64%. Figure 2 show that the FEM simulated analysis and experimental results are generally in agreement. It is noted from figure 2 that there is an difference in elastic modulus for the specimen exposed to hydrogen in the experimental data presented. This differs from model results which assumes that elastic properties are unchanged by hydrogen. Plastic work is spent solely in the generation and migration of dislocation, so hydrogen (which affects only the plastic component) will not change the elastic constants. In experiments, a small disruption in lattice structural and elastic constants may occur as observed by Lee [29] so it is inferred that this change is reflected in the observed discrepancy in the elastic part of the deformation. The current study seeks to understand the effect of hydrogen on plastic deformation and fracture vis-a-vis the HELP mechanism using the large strain theory. Plastic deformation is much larger in comparison to this discrepancy so these minor excursions in elasticity have been ignored but will be included in future model. Model parameters used are presented in Table 1.

Table 1 Model validation parameters

| Model Parameters | $H_i$ | $C_{INITIAL}$(%at) | $K_1\sqrt{y}$ | $N_L/K_T$ | $c_{Total}$ (%at) | $H_f$ |
|---|---|---|---|---|---|---|
| | 1.6 | 0.05 | 2.4 | 5.5e+26 | 0.64 | 0.05 |

$C_{initial}$ is calculated from relationship in (27). $N_L$ is a material constant given by $N_A/V_M$. $N_A$ is Avogadro's number

and $V_M$ is molar volume. $K_T$ is the equilibrium constant given by $\exp\left(\frac{W_b}{RT}\right)$ $W_b$ is binding energy, R is universal gas constant and T is absolute temperature. $C_{Total}$ is selected to match experimental data used for validation. $K_1\sqrt{y}$ is a material constant that captures athermal dislocation multiplication with respect to straining during stages 1 and 2. $K_1$ is a constant associated with athermal storage of moving dislocations which become immobilised after traveling a distance proportional to the average spacing between dislocations, $\sqrt{y}$. $H_i$ and $H_f$ are selected to match experimental results for hydrogen charged samples- both terms are zero in a material not exposed to hydrogen. To investigate the effects of hydrogen and other key parameters, a parametric review has been performed. $H_i$, $H_f$ $C_{initial}$, $C_{Total}$, $k_1\sqrt{y}$ and other parameters have been varied over selected ranges to observe the individual effects on the material stress – strain response. The stress - strain relationship of the material with different values of $H_i$ and $H_f$ are presented in Figure 3 and Figure 4 respectively. Parameters used for these assessments are presented in Table 2 and Table 3

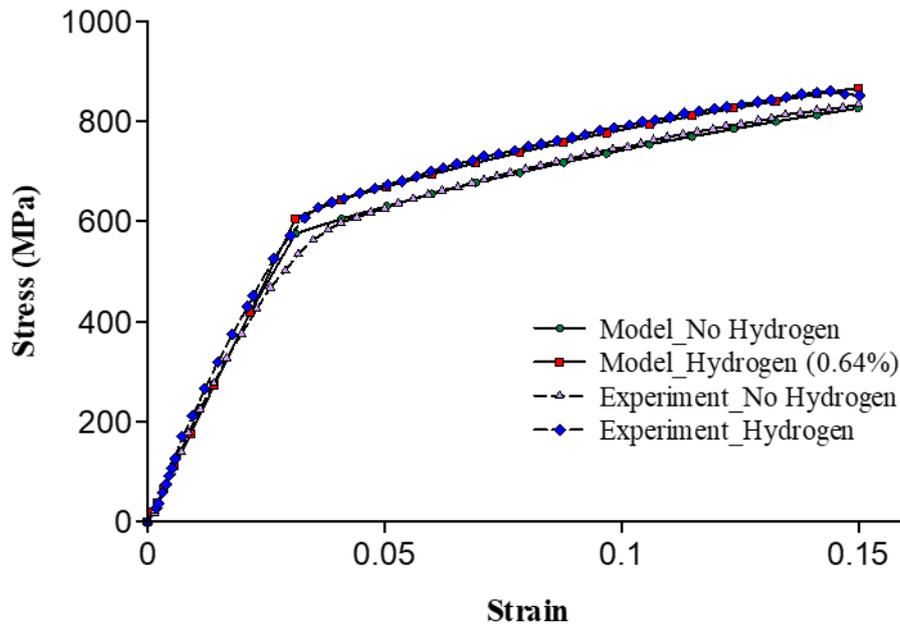

Figure 2  Simulated results with experiments data (Yagodzinskyy 2014)

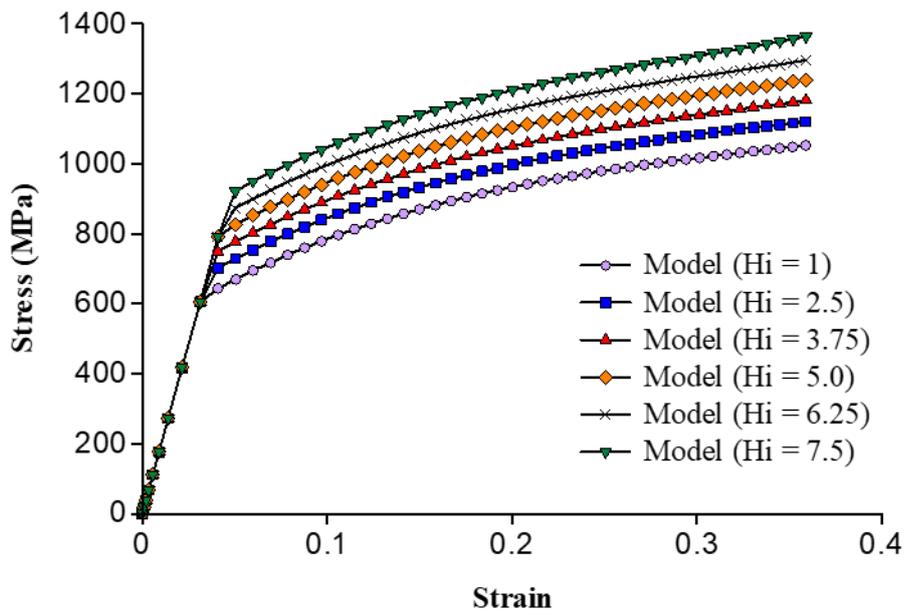

Figure 3  Stress strain curves with varying Hi

Results observed were consistent with experimental observations, the initial critical resolved shear stress in-creased with $H_i$ (see Figure 3). There was also a shift in the graph to higher stresses required for plastic deformation to occur representing the "locking" effect hydrogen has on Frank-Read and other sources of dislocation. Material work hardening was observed to increase with $H_f$ (see Figure 4). The increase in slope of the curve represents the ageing effect of dislocation by hydrogen atoms in austenitic stainless steel grains. Figure 5 shows the variance in the stress strain curve with hydrogen content. It can be observed that there was an increase in uniaxial tension yield strength of the austenitic stainless steel single crystals as reported by experiments performed by Yagodzinskyy et al [30]. There is also a visible increase in work hardening as observed by Delafosse [31].

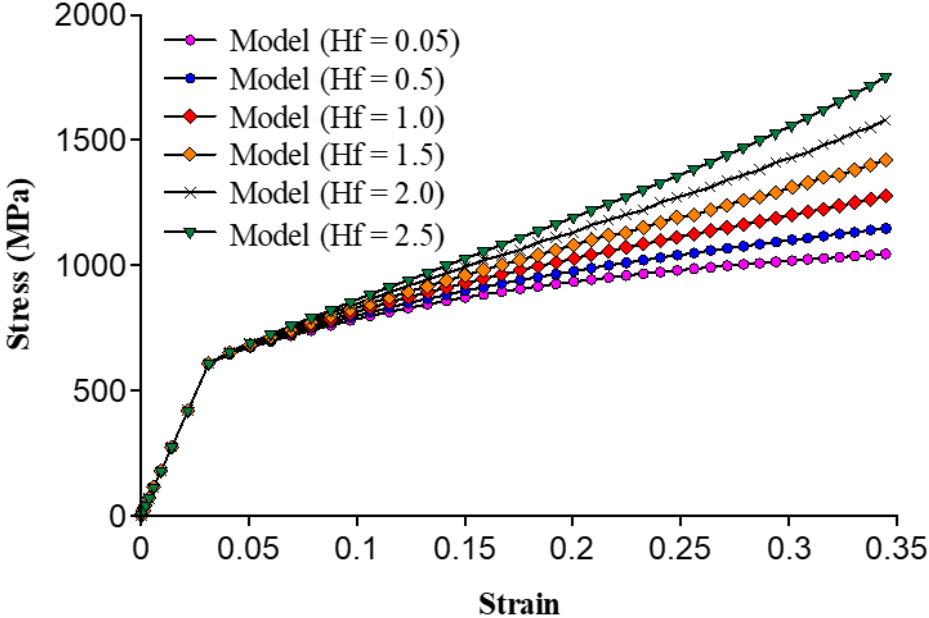

Figure 4  Stress strain curves with varying Hf

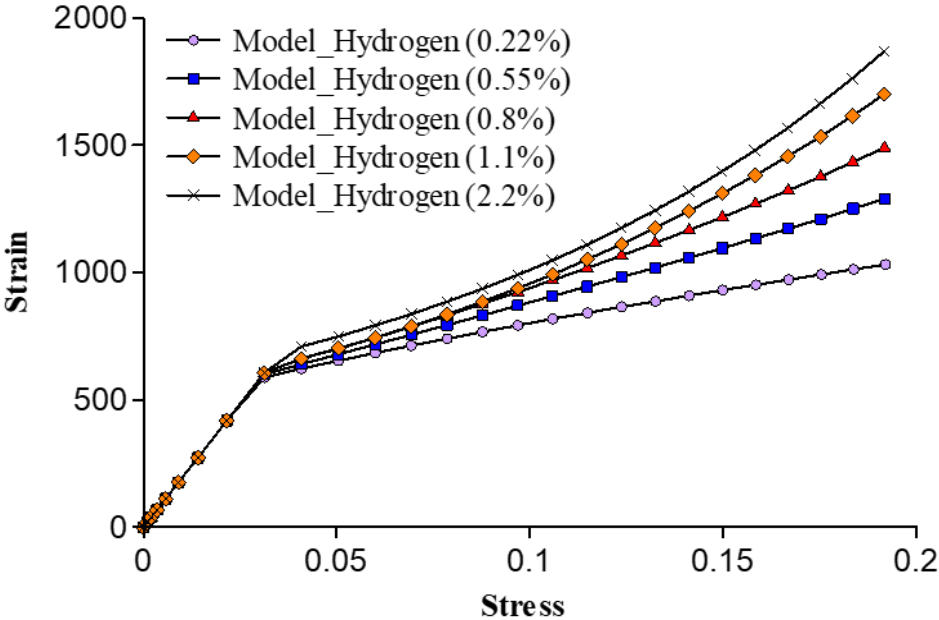

Figure 5  Stress strain curves with varying Hydrogen content

Table 2 Model validation parameters (for $H_i$)

| Model Parameters | $H_i$ | $C_{INITIAL}$(%at) | $K_1\sqrt{y}$ | $N_L/K_T$ | $c_{Total}$ (%at) | $H_f$ |
|---|---|---|---|---|---|---|
| | 1, 2.5, 3.75, 5.0, 6.25, 7.5 | 0.05 | 2.4 | 5.5e+26 | 0.64% | 0.05 |

Table 3 Model validation parameters (for $H_f$)

| Model Parameters | $H_i$ | $C_{INITIAL}$(%at) | $K_1\sqrt{y}$ | $N_L/K_T$ | $c_{Total}$ (%at) | $H_f$ |
|---|---|---|---|---|---|---|
| | 10 | 0.05 | 2.4 | 5.5e+26 | 0.64% | 0.05, 0.5, 1.0, 1.5, 2.0, 2.5 |

For the void growth analysis, an RVE model with an embedded void of known initial void fraction is constructed and meshed using ABAQUS/CAE with reduced-integration elements (C3D8R) (see illustration of void model in Figure 6).

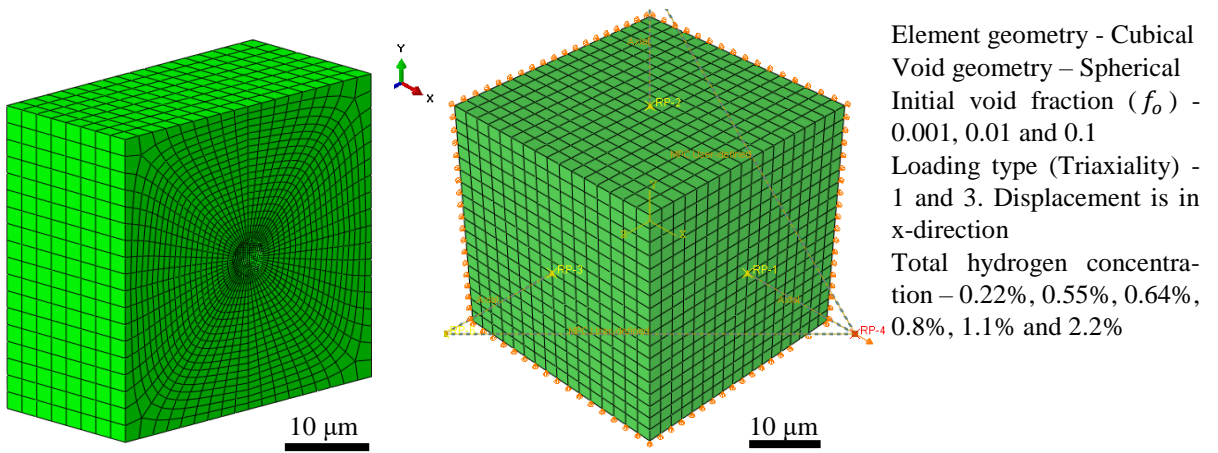

Element geometry - Cubical
Void geometry – Spherical
Initial void fraction ($f_o$) - 0.001, 0.01 and 0.1
Loading type (Triaxiality) - 1 and 3. Displacement is in x-direction
Total hydrogen concentration – 0.22%, 0.55%, 0.64%, 0.8%, 1.1% and 2.2%

Figure 6   RVE Specimen half sectioned to show spherical embedded void

The porous crystal plasticity model and the relationship between void growth, strain, stress triaxiality, initial void size and crystal orientation have been discussed by other authors [32, 33, 34] so only a summary is given here. Void fraction evolution is defined as follows;

$$f_o = \frac{V_{void}}{V_{Total}} \qquad (29)$$

$V_{void}$ is the void volume and $V_{total}$ is the total volume of the element i.e. solid material and void. Using the validated material parameters from Table 1, void growth analyses have been performed over selected ranges of hydrogen to observe the effects on the material stress strain response and analyse the effect of hydrogen on void growth. Displacements in the lateral direction were tuned to keep applied stress triaxialities constant using a multipoint constraint (MPC) user subroutine of the ABAQUS software, while volume averaged stress triaxiality was varied depending on the void growth. Figure 7 shows that for the RVE embedded with a void, hydrogen was observed to increase the overall equivalent plastic stresses for similar equivalent strain values. This indicates a hardening effect of atomic hydrogen. Results also show higher values of void fraction at stress triaxiality of 3 when compared to stress triaxiality of 1 (Figure 8). This was observed for both hydrogen containing and hydrogen free conditions. Atomic hydrogen is inferred to have slowed void growth for both triaxialities 1 and 3 as lower void fraction values were observed for hydrogenated element for similar equivalent strain values. This observation is consistent with experimental findings for austenitic stainless steels and other FCC metals exposed to hydrogen [35, 36]. Hydrogen in traps concentrate around the boundary of the embedded void during elastic deformation and there is little hydrogen concentration in traps observed in regions remote from the void (Figure 9a & 9b). As equivalent stress values approach yield, hydrogen transfer becomes more intense in areas in close proximity to the void (Figure 9c) and then extends to regions which are remote from the void (See Figure 9d). This progression in hydrogen exchange from NILS to trapping sites is induced by increased deformation at these sites and is consistent with observations in literature [37]. As noted in the introduction section, experimental observations have shown that hydrogen atoms act to "pin" down dislocations groups [8]. It can be inferred from results that hydrogen reduces void growth by immobilising dislocations immediately bounding the void. This leads to an increase in the work hardening and slowdown in dislocation motion in the material region which would restrict void expansion. This phenomenon has not been observed in the hydrogen free simulations where a higher values of void fraction have been reported for similar equivalent strains values.

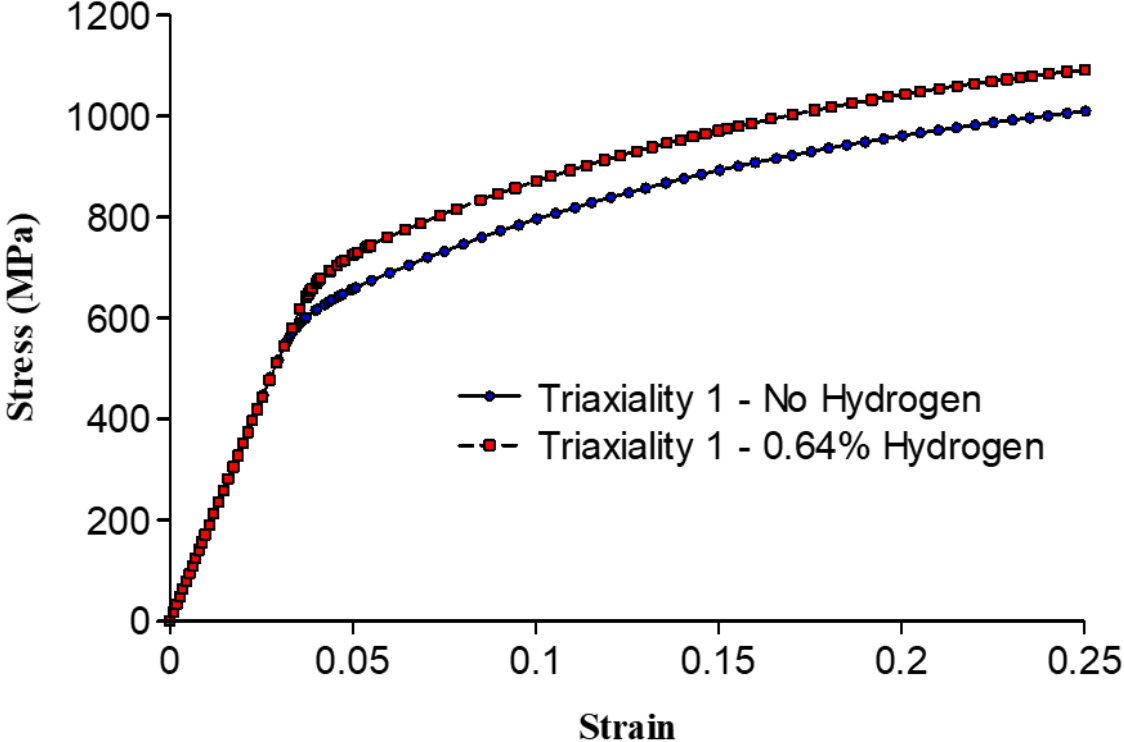

Figure 7  Stress vs Strain curves at Stress Triaxiality = 1

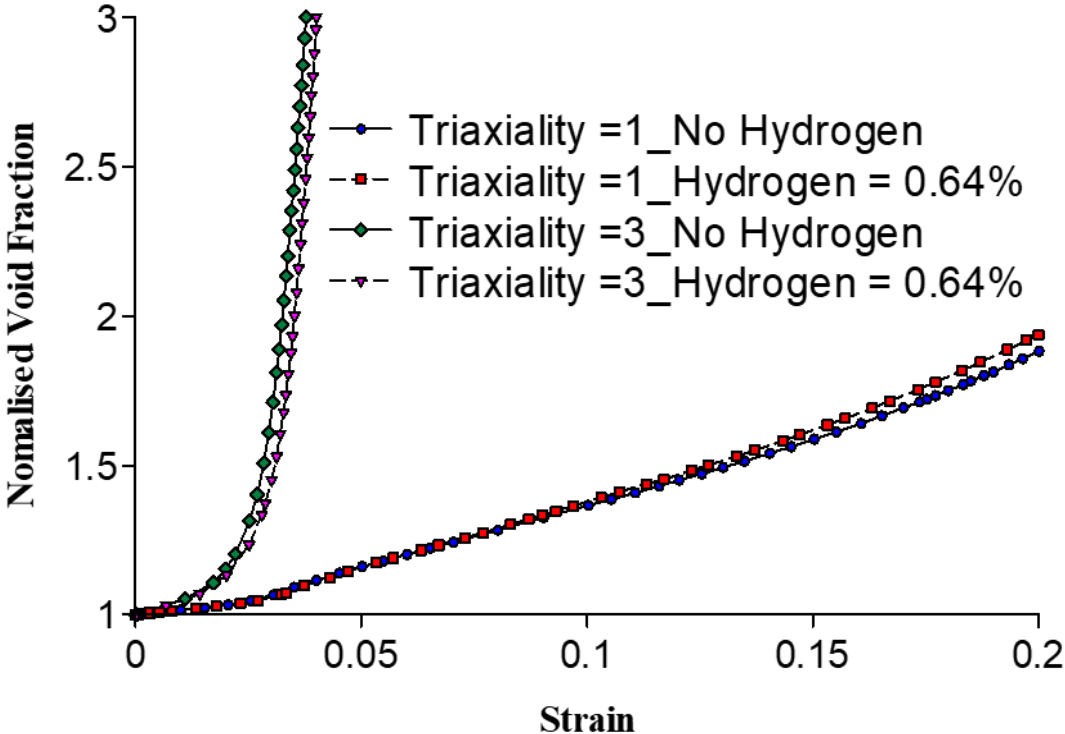

Figure 8  Normalised void fraction relationship with equivalent strain

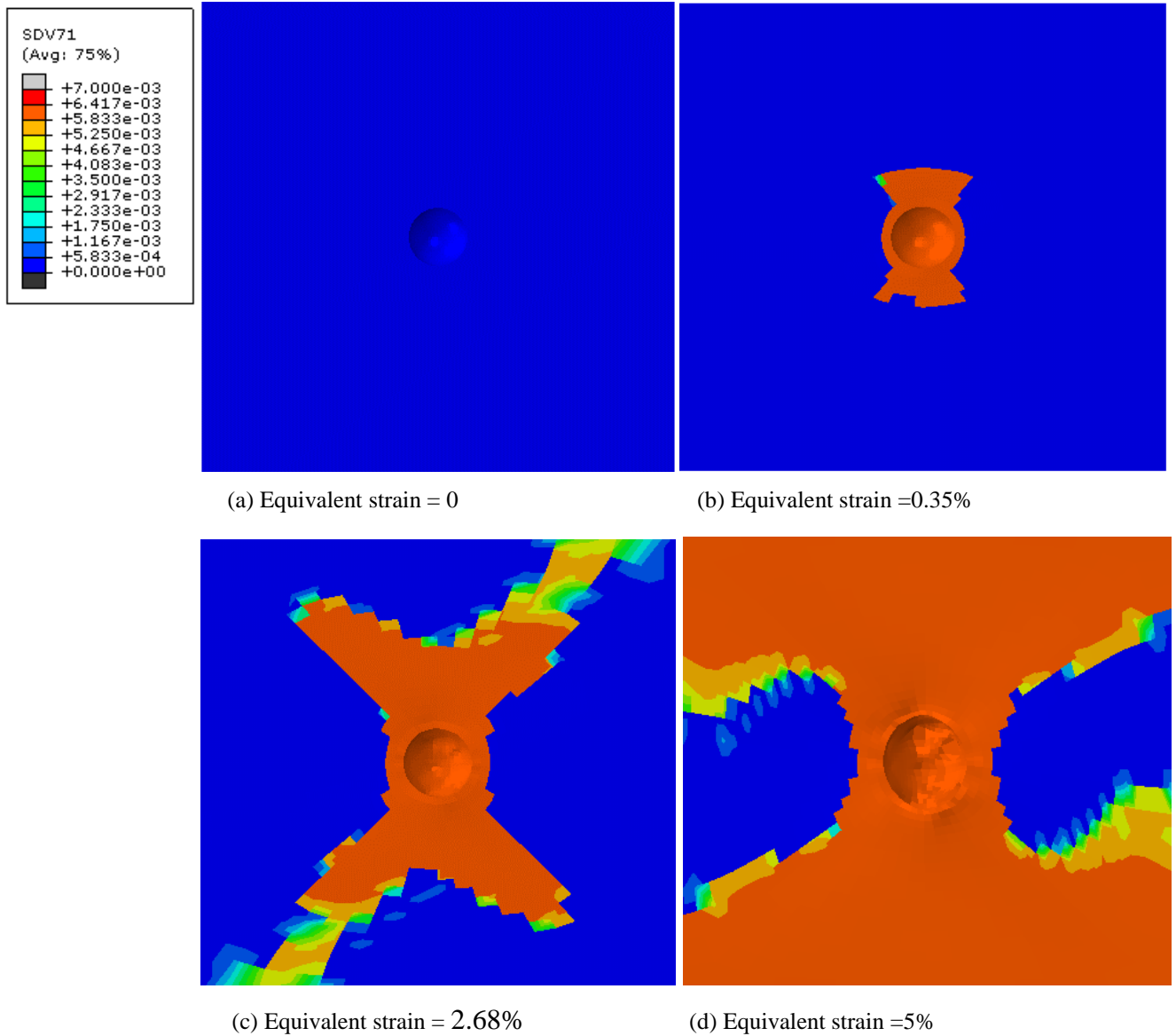

(a) Equivalent strain = 0  (b) Equivalent strain = 0.35%

(c) Equivalent strain = 2.68%  (d) Equivalent strain = 5%

Figure 9   Concentration of hydrogen in trapping sites expressed in % atom (SDV 71) (a) No strain applied (b) During elastic deformation (c) Early stages of plastic deformation (d) Later stages of plastic deformation

## 5. CONCLUSIONS

The model presented simulates the hydrogen induced stress corrosion cracking mechanism by capturing experimental knowledge of hydrogen –dislocation interactions within a crystal plasticity computational framework. Hydrogen concentration in traps have been modelled to increase the critical resolved shear stress and work hardening of the material. These have been captured by introducing two terms. The hydrogen initial strength coefficient $H_i$ quantifies the effect of hydrogen on the initial yield strength of the crystal. Hydrogen in traps increase dislocations density which leads to macroscopic crystal hardening and this has been captured by a term called hydrogen hardness coefficient ($H_f$). Void growth is slowed by hydrogen which accumulate around the void boundary during the elastic part but its effects extend to areas remote from the void as plastic deformation progresses. Higher critical resolved stress values were observed in the presence of hydrogen for various stress triaxialities for both voided and non-voided RVE samples. Hydrogen was found to reduce void growth by immobilising dislocations leading to an increase in the work hardening. Experimental results have shown that an increase in void density and coalescence is favoured instead of void growth in FCC structured austenitic stainless steels in the presence of hydrogen [8, 16, 38]. This phenomenon is typically characterised by a reduction in the average size of dimples on fracture surfaces observed after failure of austenitic stainless steel [35]. Hydrogen effects on void density have not be captured in this model but is subject to future consideration and future work will extend the model to capture the effect of hydrogen on void coalescence.

**ACKNOWLEDGEMENTS**

The authors are thankful to the University of Aberdeen and Apache North Sea for their support for this project.

**REFERENCES**

1. McGuire, M.F., "Stainless Steels for Design Engineers", ASM International, pp 296 (2008).
2. Painkra, T.K., Naik, K.S., Nishad, R.K., Sen, P.K., "SK. Review about high performance of austenitic stainless steel," *International Journal for Innovative Research in Science & Technology*, pp. 93-99 (2014).
3. Woodtli, J., Kieselbach, R., "Damage due to hydrogen embrittlement and stress corrosion cracking," *Engineering Failure Analysis*, pp. 427-450 (2000).
4. Birnbaum, H.K., Sofronis, P., "Hydrogen-enhanced localized plasticity—a mechanism for hydrogen-related fracture," *Materials Science and Engineering*, volume 176, issues 1-2, pp.191-202 (1994).
5. Abraham, D.P., Altstetter, C.J., "The effect of hydrogen on the yield and flow stress of an austenitic stainless steel," *Metallurgical and Materials Transactions A*, volume 26, issue 11, pp. 2849-2858 (1995).
6. Robertson, I.M., "The effect of hydrogen on dislocation dynamics," *Engineering Fracture Mechanics*, volume 64, issue 54, pp. 649-673 (1999).
7. Sofronis, P., Birnbaum, H.K., "Mechanics of the hydrogen dislocation impurity interactions - I. Increasing shear modulus," *Journal of the Mechanics and Physics of Solids*, volume 43, issue 1, pp. 49-90 (1995).
8. Yagodzinskyy, Y., Malitckii, E., Saukkonen, T., Hänninen, H., "Hydrogen-induced strain localization in austenitic stainless steels and possible origins of their hydrogen embrittlement," *Proceedings of the Steel and Hydrogen 2nd International Conference on Metals and Hydrogen*, Gent, Belgium, (May 2014).
9. Skipper, C., Leisk, G., Saigal, D., Matson., "Effect of Internal Hydrogen on Fatigue Strength of Type 316 Stainless Steel," *Effects of hydrogen on materials- proceedings of the 2008 international hydrogen conference*, Wyoming, USA (September 7-10, 2008).
10. Yagodzinskyy, Y., Saukkonen, T., Tuomisto, F., Hänninen, H., "Effect of Hydrogen on Plastic Strain Localization in Single Crystals of Nickel and Austenitic Stainless Steel," *Effects of hydrogen on materials- proceedings of the 2008 international hydrogen conference*, Wyoming, USA (September 7-10, 2008).
11. Schebler, G., J., "On the mechanics of the hydrogen interaction with single crystal plasticity," M. S. Thesis, Mechanical Science & Engineering, University of Illinois at Urbana, Champaign, USA (2011).
12. Lynch, S., "Progress towards understanding mechanisms of hydrogen embrittlement and stress corrosion cracking," *proceedings of NACE International CORROSION Conference*, Tennessee, USA (March 11-15, 2007)
13. Takai, K., Shoda, H., Suzuki, H., Nagumo, M., "Lattice defects dominating hydrogen-related failure of metals," *Acta Materialia*, volume 56, issue 18, pp.5158-5167 (2008).
14. Martin, M.L., Robertson, I.M., Sofronis, P., "Interpreting hydrogen-induced fracture surfaces in terms of deformation processes: A new approach," *Acta Materialia,* volume 59, issue 9, pp.3680-3687 (2011).
15. Martin, M.L., Fenske, J.A., Liu, G.S., Sofronis P., Robertson, I.M., "On the formation and nature of quasi-cleavage fracture surfaces in hydrogen embrittled steels," *Acta Materialia,* volume 59, issue 4, pp.1601-1606 (2011).
16. Bullen, D., Kulcinski, G., Dodd, R., "Effect of hydrogen on void production in nickel," *Journal of Nuclear Materials*, volumes 133-134, pp.455-458 (1985).
17. Matsumoto, Y., Kurihara, N., Suzuki, H., Takai, K., "Hydrogen embrittlement and hydrogen-enhanced strain-induced vacancies in α-iron," *TMS 2017 146th Annual Meeting & Exhibition Supplemental Proceedings*, California, USA (2017)
18. Cuitino, A., Ortiz, M., "Ductile fracture by vacancy condensation in FCC single crystals," *Acta Materialia,* volume 44, issue 2, pp.427-436 (1996).
19. Hill, R., Rice, J., "Constitutive analysis of elastic-plastic crystals at arbitrary strain," *Journal of the Mechanics and Physics of Solids*, volume 20, pp. 401-413 (1972).
20. Marin, E.B., "On the formulation of a crystal plasticity model," Sandia National Laboratories, Sandia Report 2006.
21. D.M.B., Defence guide DG-8, "Treatments for protection of metal parts of service stores and equipment against corrosion," *British Corrosion Journal*, pp.121 (2013).
22. Oriani, R., "Hydrogen embrittlement of steels," *Annual Review of Materials Science*, pp 327-357 (1978).
23. Estrin, Y., Mecking, H., "A unified phenomenological description of work hardening and creep based on one-parameter models," *Acta Materialia,* volume 32, issue 1, pp.57-70 (1984).

24. Kocks, U.F., Mecking, H., "Physics and phenomenology of strain hardening: the FCC case," *Progress in Materials Science*, volume 48, issue 3, pp 171-273 (2003).
25. Somerday, B., Dadfarnia, M., Balch, D., Nibur, K., Cadden, C., Sofronis, P., "Hydrogen-assisted crack propagation in austenitic stainless steel fusion welds," *Metallurgical and Materials Transactions*, volume 40, issue 10, pp 2350-2362 (2009).
26. Krom, A.H.M., "Numerical modelling of hydrogen transport in steel" ," Doctoral Thesis, Applied Sciences, Delft University of Technology, Delft, Netherlands (1998).
27. Caskey, G.R. Jr, "Hydrogen solubility in austenitic stainless steels," *Scripta Metallurgica*, volume 34, issue 2, pp 1187-1190 (1981).
28. Dassault Systèmes Simulia Corp, "ABAQUS 6.14 documentation", Dassault Systemes, Providence, RI, USA (2014).
29. Lee, E.H., "Elastic-Plastic Deformation at Finite Strains," *Journal of Applied Mechanics*, volume 36, issue 1, pp 1-6 (1969).
30. Yagodzinskyy, Y., Tarasenko, O., "Effect of hydrogen on plastic deformation of stable 18Cr-16Ni-10Mn austenitic stainless steel single crystal," *Effects of hydrogen on materials- proceedings of the 2008 international hydrogen conference*, Wyoming, USA (September 7-10, 2008).
31. Delafosse, D., Y., Feaugas, X., Aubertc, I., Saintier, N., Olive, J.M., "Hydrogen effects on the plasticity of fcc nickel and austenitic alloys," *Effects of hydrogen on materials- proceedings of the 2008 international hydrogen conference*, Wyoming, USA (September 7-10, 2008).
32. Asim, U., Siddiq, M.A., Demiral, M., "Void growth in high strength aluminium alloy single crystals: A CPFEM based study," *Modelling and Simulation in Materials Science and Engineering*, volume 25, issue 3, pp. 35 (2017).
33. Siddiq, A., "A porous crystal plasticity constitutive model for ductile deformation and failure in porous single crystals," *International Journal of Damage Mechanics*, volume 28 issue 2, pp. 233-248 (2018).
34. Asim, U.B., Siddiq, M.A., Kartal, M.E., "Representative volume element (RVE) based crystal plasticity study of growth on phase boundary in titanium alloys," *Computational Material Science,* volume 161, pp. 346-350 (2019).
35. Liang, Y., Ahn, D., Sofronis, P., Dodds, H., Bammann, D., "Effect of hydrogen trapping on void growth and coalescence in metals and alloys," *Mechanics of Materials*, volume 40 issue 3, pp. 115-132 ( 2008).
36. San Marchi, C., Somerday, B., Tang, X., Schiroky, G., "Effects of alloy composition and strain hardening on tensile fracture of hydrogen-precharged type 316 stainless steels," *International Journal of Hydrogen Energy*, volume 33 issue 2, pp. 889-904 (2008).
37. Ahn, D., Sofronis, P., Dodds, R., "On hydrogen-induced plastic flow localization during void growth and coalescence," *International Journal of Hydrogen Energy*, volume 32 issue 16 pp. 3734-3742 (2007).
38. Matsuo, T., Yamabe, J., Matsuoka, S., "Effects of hydrogen on tensile properties and fracture surface morphologies of type 316L stainless steel," *International Journal of Hydrogen Energy*, volume 39 issue 7 pp. 3542-3551 (2014)